# Persistent Data Layout and Infrastructure for Efficient Selective Retrieval of Event Data in ATLAS

P. van Gemmeren, D. Malon
*ANL, Lemont, IL 60439, USA*

The ATLAS detector at CERN has completed its first full year of recording collisions at 7 TeV, resulting in billions of events and petabytes of data. At these scales, physicists must have the capability to read only the data of interest to their analyses, with the importance of efficient selective access increasing as data taking continues. ATLAS has developed a sophisticated event-level metadata infrastructure and supporting I/O framework allowing event selections by explicit specification, by back navigation, and by selection queries to a TAG database via an integrated web interface. These systems and their performance have been reported on elsewhere. The ultimate success of such a system, however, depends significantly upon the efficiency of selective event retrieval. Supporting such retrieval can be challenging, as ATLAS stores its event data in column-wise orientation using ROOT trees for a number of reasons, including compression considerations, histogramming use cases, and more. For 2011 data, ATLAS will utilize new capabilities in ROOT to tune the persistent storage layout of event data, and to significantly speed up selective event reading. The new persistent layout strategy and its implications for I/O performance are described in this paper.

## 1. Introduction

The Large Hadron Collider (LHC) at CERN provides high energy proton-proton collisions to four large experiments. The ATLAS experiment [1] is one of the two multi-purpose detectors with a wide physics program. ATLAS was designed for the discovery of new physics such as finding the Higgs boson and super-symmetric particles, and to improve precision measurements of gauge bosons and heavy quark properties. Approximately 3,000 scientists work on the ATLAS experiment, bringing a diverse range of physics objectives and strategies and a wide variety of data analysis use cases.

## 2. ATLAS Software & Event Data Model

Simulation, reconstruction, and analysis are run as part of the Athena framework [2] (see Figure 1). The Athena software architecture belongs to the blackboard family: using the most current (transient) version of the event data model, data objects are posted to a common in-memory database (or "blackboard") from which other modules can access them and produce new data objects. Such an approach greatly reduces the coupling among algorithms, tools and services.

The data associated with a single collision of proton bunches is called an "event". Within an event, ATLAS arranges its data in top-level data objects and containers that are collections of objects of homogeneous type.

StoreGate [3] is the Athena implementation of the blackboard and allows a module to transparently use a data object created by an upstream module or read from disk. StoreGate uses proxies to define and hide the cache-fault mechanism: upon request, a missing data object is created and added to the transient data store, reading it from persistent storage on demand. All data retrieval from persistency is initiated by StoreGate.

StoreGate identifies objects via their data type and a string key. It supports base-class and derived-class retrieval of data objects and containers, key aliases, and inter-object references. StoreGate retrieval granularity is top-level data objects and container. StoreGate does not support partial object retrieval or strategies such as selective reading of attributes or container elements from the persistent store.

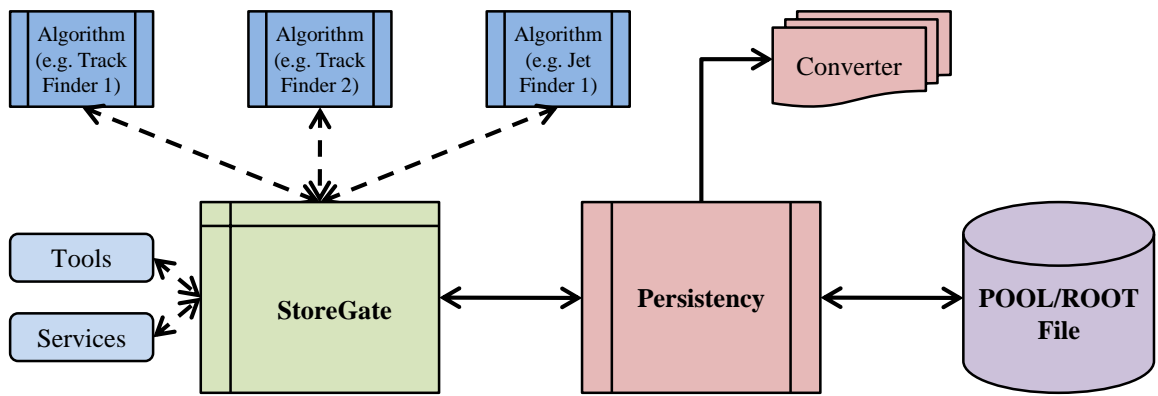

Figure 1: Layout of the ATLAS software framework.

RAW data are events as delivered by the Event Filter for reconstruction, and are essentially a serialization of detector readouts, trigger decisions and Event Filter calculations and are stored in a so-called bytestream format. The current uncompressed RAW event size is 1.6 megabytes which can be reduced by about 50% using event compression. Events are arriving at a rate of about 300 to 400 hertz. With the expected duty cycle of the ATLAS detector and the LHC, ATLAS anticipates recording more than three petabytes of RAW data per year.

Event Summary Data (ESD) refers to event data written as the output of the reconstruction of RAW data. ESD content is sufficient to render access to RAW data unnecessary for most physics applications other than calibration or re-reconstruction. ESD are stored in POOL/ROOT files, described in Section 3 below. The current size is approximately 700 kilobytes per event.

Analysis Object Data (AOD) provides a further reduced event representation, derived from ESD. Its content is suitable for physics analysis. AOD, like ESD, is stored in POOL/ROOT files. The current AOD size is just below 200 kilobytes per event on average.

Samples derived from AOD are called Derived Physics Data (DPD) and may be further streamed according to the physics selection and purpose of the samples. They too can be stored in POOL/ROOT files

## 2.1. Event Selection using TAGs

Event tags (TAG) [4] are event-level metadata which support efficient identification and selection of events of interest to a given analysis. TAG content includes, for example, trigger decisions, event properties (like missing transverse energy), and particle data (jets, electrons, muons, photons, …). To facilitate queries for event selection, TAG data can be stored in ROOT files or a relational database. The current TAG size is approximately 1 kilobyte per event.

TAGs allow jobs to process only those events that satisfy a given predicate. Less complex predicates can be applied as SQL-style queries on the TAG database using a web browser interface or on TAG files directly. For more complex criteria, C++ modules can be written using the TAG attributes for event selection. No payload data is retrieved for unselected events and data files containing only unselected events are not accessed.

## 3. POOL/ROOT Persistency Framework

ATLAS offline persistency software [5][6] uses ROOT I/O [7] via the POOL persistency framework [8], which provides high-performance and highly scalable object serialization to self-describing, schema-evolvable, random-access files (see Figure 2).

POOL software is based upon a hybrid approach and combines two technologies with different features into a single consistent API and storage system.

The first technology, object streaming software (e.g., ROOT I/O), addresses persistency for complex C++ objects such as high energy physics event data. Such data often is used in a write-once, read-many mode so that concurrent access can be constrained to the simple read-only case and no central services are needed to implement transaction or locking mechanisms. In practice, ROOT I/O is the only technology currently supported by POOL for object streaming into files.

The second technology, Relational Database (RDBMS, e.g., MySQL), provides distributed, transactionally consistent, concurrent access to data that may need to be updated. RDBMS-based stores also provide facilities for efficient server side query evaluation. RDBMS-based components are currently used in the area of catalogues, collections, and metadata; streaming technology is used for the bulk data.

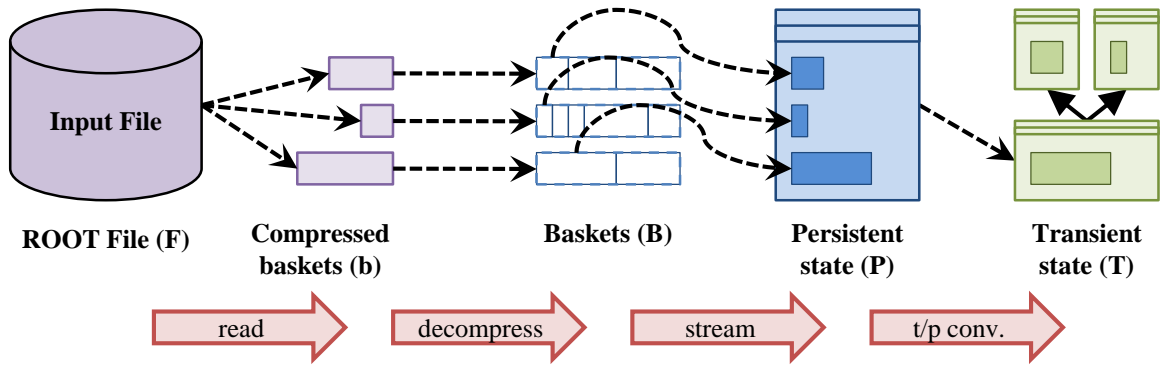

Figure 2: Reading of ATLAS data from a POOL/ROOT file.

In ROOT I/O, a file can contain directories and objects organized in unlimited number of levels, and it is stored in machine-independent format. ROOT files start with a header (100 bytes, including file identifier, version and 64-bit sizes and offsets), a top directory description (84 bytes, including name, creation and modification date) and then several logical records of variable length. The first 4 bytes of each physical record are an integer holding the size of this record followed by all the information necessary to uniquely identify a data block in the file. These records are followed by a list of class descriptions for all objects that have been written to the file.

## 4. ROOT Object Streaming & Splitting

ROOT writes data objects as trees. A ROOT tree is made of branches and each branch is described by its leaves. The leaves can be of native data types, structures, variable length arrays, or classes. The state of the objects is captured by special methods called streamers. These streamers decompose data objects into their data members. A streamer of a class object can call the streamer of all base classes, moving up the inheritance hierarchy. Also the streamers for all contained or associated objects can be called successively writing buffers that contain all data of all classes that make up a particular object. Trees can hold multiple objects of the same kind as separate entries.

The ROOT tree class is optimized to reduce disk space and enhance access speed for the persistified objects. Each branch can be read independently from any other branch as they are placed in different baskets (described below). ROOT I/O allows direct access to any entry, to any branch, or to any leaf, even in the case of variable length structures.

When ROOT writes composite objects to tree, data members can be split into separate branches. The depth is controlled by a configurable split-level. Split-level equal 0 means the whole object is written entirely to a single branch. A split-level of 1 will assign a branch to each object data member, for split-level 2, composite data members are also split and each further increase of the split-level will increase the splitting depth.

When splitting trees, each branch will have its own buffer, called a basket. When they reach a configurable basket size, these baskets can be compressed individually and are flushed to the output file. Often, a split tree has smaller disk size due to higher compression and can be faster to read, but may be slower to write and require more memory (due to the increase in buffers). One of the main advantages of splitting is the fact that each branch can be read independently, therefore increasing split-level leads to finer retrieval granularity. Data members contained via pointer cannot be split.

## 5. History of Storage Layout

ATLAS started using POOL/ROOT in 2002, when the POOL project was initiated. At that time, ATLAS wrote a separate tree for each StoreGate collection, because POOL worked correctly only for containers mapped to ROOT trees. Also, ATLAS persistified its transient data model directly without an insulation layer. Because the transient data model uses DataVector classes, which contain pointers to the data objects, for all containers, no ROOT splitting was done.

To enable arbitrary schema evolution and enhance performance and robustness of ATLAS I/O, a separate persistent data model and transient-persistent mapping layer were developed. To maximize the benefits of this transient-persistent separation, the persistent classes are as simple as possible, containing only basic C++ types. Inheritance, polymorphism and pointers are avoided and in general no methods are necessary. Because DataVector classes and their pointers were not used in the persistent data model, ROOT could split the data objects and ATLAS utilized the ROOT default of split-level of 99, causing the data objects to be fully split.

After functionality was added to POOL to allow creation of a container matched to a branch, ATLAS decided to assemble most of the event data in a single tree. The basket sizes were set to be equal for all baskets in a tree and ATLAS used the POOL default of 16 KB, until memory constraints forced us to lower it to 2 KB. In a separate processing step, output files were optimized by reordering the baskets within a file so they would be accessed more sequentially. Starting with data taken in 2011, ATLAS uses a new ROOT feature to optimize basket sizes in the main event tree[1], to share 30 MB so that all baskets have approximately the same number of entries. At this point, basket reordering was deemed to be no longer necessary.

---

[1] Basket optimization is only used for the main event tree, which contains more than 90% of the data.

## 6. Mismatch of Transient vs. Persistent Event Data Layout

In HEP, event data associated with individual collisions are processed separately. Therefore the ATLAS transient data store, StoreGate, is event-wise oriented, holding just the data for the event being processed. Once the processing is done, all event data is cleared from StoreGate and the data for the next event is made accessible for reading. StoreGate allows the on-demand retrieval of top-level data objects and container. Once an object or container is requested from the transient store, StoreGate cause the persistency infrastructure to read it completely. Partial object retrieval (e.g., reading only some of the data members or a fraction of container elements from persistent storage) is not currently supported by StoreGate. The retrieval granularity for StoreGate corresponds to top-level objects and to containers.

The persistent event data layout is different. ROOT trees are stored column-wise and the width of the column depends on the split level. For fully split trees each column holds an individual data member of built-in type. The data from each column is stored in individual baskets, containing multiple entries, which are compressed to save disk space.

When retrieving a data object for a single event, the entire baskets for all the individual data members have to be read and decompressed. Most of these baskets contain data for the following (or preceding) events (see Figure 3). If the data object is not retrieved for these events, than the I/O and CPU effort was wasted.

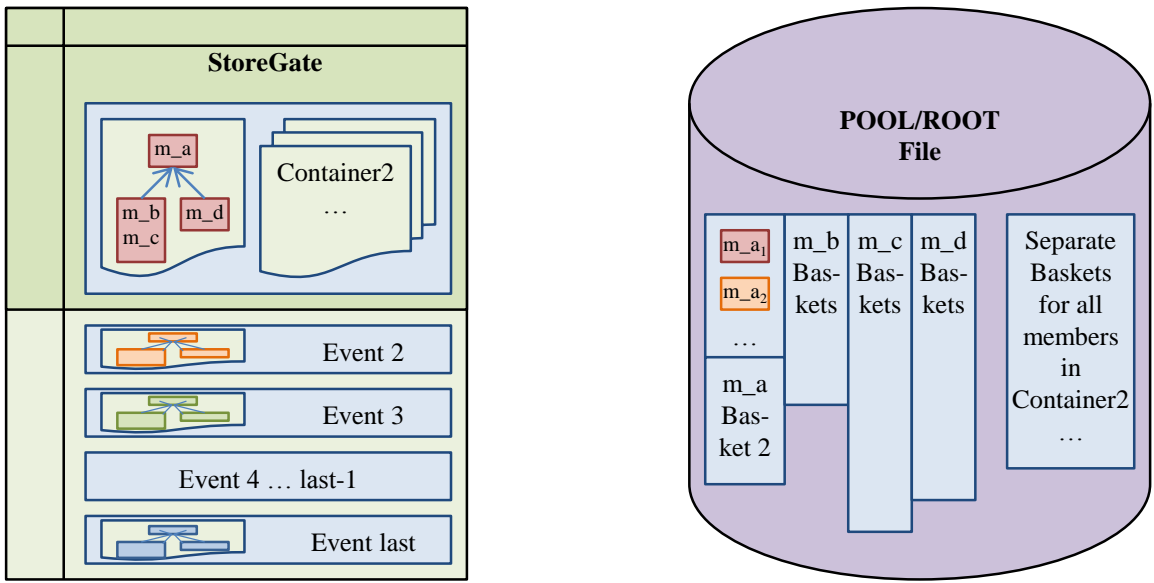

Figure 3: Mismatch of ATLAS transient StoreGate versus persistent event data layout in POOL/ROOT files.

## 7. Changes to the Persistent Event Data Layout

The following changes were made to achieve a better match between transient data retrieval and persistent data layout. These changes leverage some recent ROOT improvements and consolidate some artifacts caused by the history of POOL/ROOT and ATLAS I/O software development.

### 7.1. No single attribute retrieval

As StoreGate does not support partial data object retrieval, ATLAS cannot take advantage of ROOT's member-wise reading capability provided by the fully split persistent data. Other advantages of splitting (e.g., compression, read speed) can be achieved to a similar degree when using member-wise streaming. On the other hand, fully splitting the complex ATLAS event data model creates a large number of baskets, approximately 10,000 each for ESD and AOD, requiring many disk reads for object retrieval (see Figure 4).

Therefore, the new event data layout switches off splitting, while continuing member-wise streaming, so each StoreGate collection (i.e., POOL container) will be stored in a single ROOT basket. This change is made for all but one of the largest container in the ESD (the Track container) and the two largest containers in the AOD (the TrackParticle and CaloCluster containers). Because of their large size, these containers show good performance with the current persistent layout and not splitting them would cause a significant file size increase. The capability to set the ROOT split-level for individual POOL container was developed recently as part of the ATLAS event data layout studies.

This change decreases the number of baskets in the ESD and AOD to about 800, and therefore increases their average size by more than a factor of 10, which lowers the number of disk reads. Using member-wise streaming and single baskets for individual data objects, is also beneficial for the basket ordering.

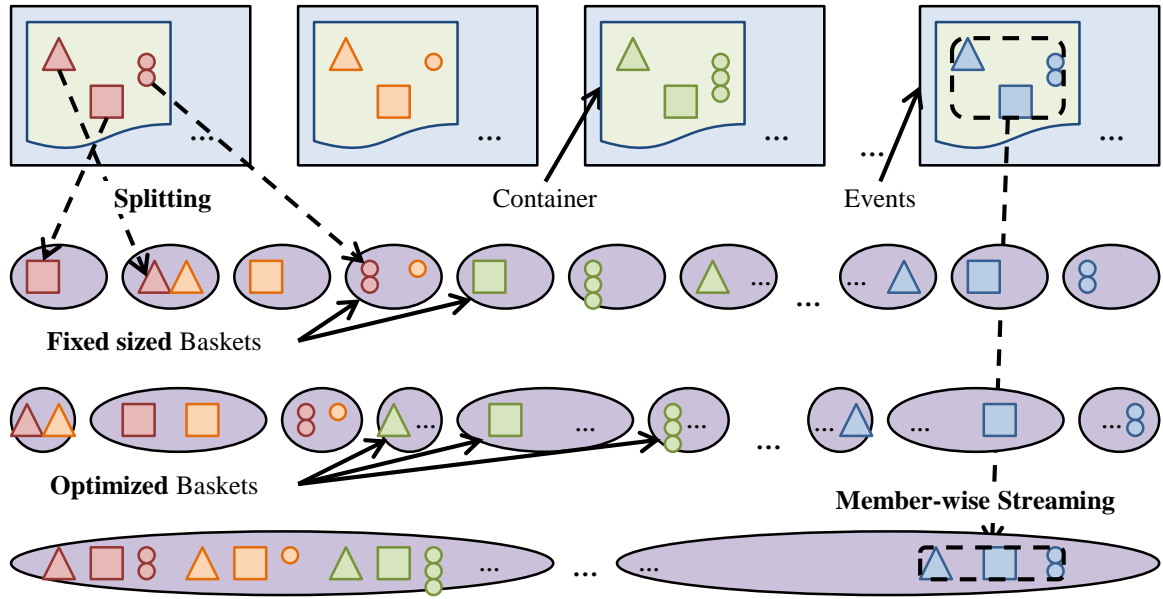

Figure 4: Different ROOT streaming modes.

## 7.2. Almost event-level retrieval

With basket sizes optimized for 30 MB ROOT trees, approximately 50 to 200 entries (i.e., events) share the same basket. For sequential reading of all data objects in every event, this does not pose a problem, but for selective reading of a single event, this means that 30 MB of data have to be read from disk and uncompressed, even when only a small fraction is needed. Before using basket size optimization, the situation was similar: For fixed size 2 KB baskets[2], the number of entries varies to a large degree. To read a single event, more than 10,000 baskets containing between approximately 5 and 10 MB of data have to be read from disk and decompressed. This behavior causes a large performance penalty on selective event retrieval.

The new data layout is designed to minimize the performance penalty for selective event reading, without hurting overall I/O performance. The change outlined in Section 7.1 reduces the number of disk reads by more than an order of magnitude, while increasing their size accordingly. This allows ATLAS to store fewer events per basket. In the new persistent data layout, automatic basket optimization is used to flush baskets every 5 events for ESD and every 10 events for AOD. The reduction in total basket size within the main event tree and switching off splitting for all trees allow ATLAS to reset the basket sizes for the other trees from 2 KB to 32 KB, which is the ROOT default.

## 8. Performance Measurements

Two scenarios were considered to test read speed performance of the new data layout. First, reading all data objects of all events in sequence, in a way that is usually done during data processing campaigns, was studied on ESD and AOD. Analysis users of the AOD often do not require all information from all events, but may read a sparse selection of events or may retrieve a particular data object only for events with certain characteristics. The second use case is studied utilizing TAGs as a mechanism to randomly read 1% of all events. For the selected events, all data objects are read.

---

[2] The 2 KB basket size is for uncompressed data, with a typical compression factor of 3, compressed baskets are between 0.5 and 1 KB. 10,000 baskets therefore compress to about 5 – 10 MB.

## 8.1. Reading all events sequentially

Even though this use case was not the main motivation for changing the persistent event data layout, it is the most common one for ESD and AOD data. It is considered mandatory that any change at least preserves the performance for sequential reading of all events, regardless of any benefits for other use cases.

Therefore sequential reading has been studied using two large samples each of ESD and AOD data, written with the same data content and software version, one using full split mode and a 30 MB optimized main event tree and the other with the new persistent event data layout.

The results of a CPU time measurement for reading all events sequentially can be found in Table I. They show that the new data layout not only matches the performance of the previous format, but improves it by approximately 15% for ESD and over 30% for AOD. The reason for the improvement is the smaller number of required disk reads combined with the better sequencing of the baskets in the file.

Table I: CPU times[3] for reading ESD and AOD events sequentially

| Reading of all events sequentially | ESD | AOD |
|---|---|---|
| Full splitting and 30 MB to optimize main event tree | 420 ms/event | 55 (±3) ms/event |
| No splitting and flushing of main event tree every 10/5 events | 360 ms/event | 35 (±2) ms/event |

The largest containers in ESD and AOD remain fully split and even with the 30 MB optimized event tree were flushed to file more frequently than smaller containers[4], so the effective changes to these containers were less significant, and therefore they do not gain or lose in I/O performance. For larger containers, I/O performance is good in either storage layout. Very small data objects often have only few data members, so there are only a few baskets even in full split mode. The small effects of reducing the number of branches by switching off splitting and flushing them to file more frequently almost cancel each other, and there is little I/O performance gain. The main contribution in I/O performance gain for sequential reading of all events comes from medium sized containers (e.g., Vertex, TrkSegment, EGamma, Muon, and Tau container gain a factor of ~ 3; the smaller Track, TrackParticle and CaloCluster container gain a factor of 3 – 7).

## 8.2. Selective reading 1% of events

The main motivation for the layout changes is the use case of selective reading. TAG files were created for the AOD files used for the tests described in section 8.1. These TAGs contain a random-valued attribute that was used to select 1% of the events. The TAG infrastructure poses only a very negligible overhead for event processing. However, when reading only a subsample of the events, a large per-event penalty is caused by the fact that the ROOT baskets contain data for many more events that need to be uncompressed, but may not be part of any requested event.

The per-event read speed times for selective reading 1% of events for AOD is given in Table II, as is the penalty caused mainly by reading and uncompressing too much data. With full splitting and 30 MB event tree, the selective reading slows down to 270 ms/event versus 55 ms/event when reading sequentially. Even though this is a very large penalty, the total read time for the job is about 20 times faster. The reason for the large increase in per-event reading time is that even for a 1% selection most of the data will be read and decompressed.

---

[3] Performance measurements, such as CPU times depend on various circumstances, such as CPU type and users of the computer. All tests were executed individually on the same machine, which was not used by anybody else. The measurements were found to be reproducible on the same machine and the relative differences between the various times were independently verified on different architectures.

[4] Even though basket optimization determines basket sizes to hold approximately the same number of entries, there is a strong bias to allow fewer entries for the largest and more entries for the smallest branches in a tree.

The smaller number of entries per basket of the new layout speeds up selective reading by a factor of 4 – 5 to just 60 ms/event and reduces the penalty caused by reading adjacent events to about 70%. The read time of a job selecting 1% of the events is almost 60 times faster than a job reading all events. In fact the selective per-event read time with the new data layout is nearly as fast as the sequential per-event read time for the old format. Providing read performance for a selected subset of events that is similar to the sequential read performance of an extracted, materialized subset may reduce the number of extracted samples produced by the collaboration.

Table II: CPU times for selective reading 1% of AOD events

| Selective reading of events. | 1% AOD | Event penalty |
|---|---|---|
| Full splitting and 30 MB to optimize main event tree | 270 ms/event | ~400% |
| No splitting and flushing of main event tree every 10/5 events | 60 ms/event | ~70% |

When reading events data selectively via TAGs, all of the data objects show great improvements in read speed. These improvements will be seen for any mechanism that accesses events or individual data objects sparsely. This will be important for AthenaMP [9], which uses multiple workers on different computing cores to read and process event data. Each worker therefore reads events only from a non-sequential part of the input file

### 8.3. Further Performance Results

Read speed is a very important metric for I/O performance, but not the only one. File size, virtual memory footprint and write speed measurements are important and performance degradation in these metrics needs to be avoided. For ESD and AOD files written in member-wise streaming mode, flushing every 5/10 events leads to no significant file size increase. The virtual memory footprint with the new storage layout is reduced by 50 – 100 MB for each reading and writing stage, caused by the fact that fewer baskets have to be loaded into memory. The write speed for ESD and AOD is increased by about 20% with the new format. For ESD the compression level can be somewhat relaxed, resulting in a write speed improvement of almost 50%.

## 9. Fall 2011 Reprocessing

In August 2011, ATLAS began a campaign to reprocess all data taken in 2011 using Release 17 software, which deploys the new storage layout. By the end of the reprocessing campaign, all 2011 data will have been reprocessed using the new storage layout and I/O performance benefits will be available to reader of ESD and AOD.

## 10. Summary

ATLAS made changes to the event data storage layout, using no ROOT splitting, but member-wise streaming and optimizing ROOT baskets to store small numbers of events. These changes cause the persistent event data layout to better match the transient access pattern and increase I/O performance. Reading all events in sequence is about 30% faster. Selective event reading speed improves even more drastically and for a 1% selection, the new format is about 4 – 5 times faster. These changes are deployed in the ATLAS fall 2011 reprocessing.

### Acknowledgments

Notice: The submitted manuscript has been created by UChicago Argonne, LLC, Operator of Argonne National Laboratory ("Argonne"). Argonne, a U.S. Department of Energy Office of Science laboratory, is operated under Contract No. DE-AC02-06CH11357. The U.S. Government retains for itself, and others acting on its behalf, a paid-up nonexclusive, irrevocable worldwide license in said article to reproduce, prepare derivative works, distribute copies to the public, and perform publicly and display publicly, by or on behalf of the Government.